\documentstyle[psfig,floats,aps]{revtex}
\begin{document}
\newcommand{\BE}{\begin{equation}}
\newcommand{\EE}{\end{equation}}
\newcommand{\rr}{{\bf r}}
\newcommand{\qq}{{\bf q}}
\newcommand{\vv}{{\bf v}}
\twocolumn
[\hsize\textwidth\columnwidth\hsize\csname @twocolumnfalse\endcsname
\draft
\title{Molecular Dynamics Simulation of Spinodal Decomposition
in Three-Dimensional Binary Fluids}

\author{Mohamed Laradji$^{(a)}$, S{\o}ren Toxvaerd$^{(b)}$
and Ole G. Mouritsen$^{(c)}$}

\address{
$^{(a)}$ Department of Physics, University of Toronto, Toronto, Ontario,
Canada M5S 1A7\\
$^{(b)}$Department of Chemistry, H. C. {\O}rsted Institute,
University of Copenhagen,\\
DK-2100 Copenhagen, Denmark\\
$^{(c)}$ Department of Physical Chemistry, The Technical University of
Denmark, Building 206, DK-2800 Lyngby, Denmark
}
\date{\today}
\maketitle

\widetext
\begin{abstract}
Using large-scale molecular dynamics simulations of a two-component Lennard-Jones
model in three dimensions, we show that the late-time
dynamics of spinodal decomposition in concentrated binary fluids 
reaches a viscous scaling regime with a growth exponent $n=1$, 
in agreement with experiments and a theoretical analysis for viscous growth.
\end{abstract}

\narrowtext

\vskip2pc]

\newpage
The dynamics of phase separation in multicomponent fluids
involves very rich and general phenomena, and has therefore been the subject
of intensive studies in recent years. The dynamics of first-order phase
transitions in general, besides being of technological importance,
is particularly interesting because of the emergence of
one characteristic length scale, $R(t)$, during the late times of the dynamics.
$R(t)$ is related to the average domain size of the ordering phase, and displays
a simple power-law dependence with time, $R(t)\sim t^n$,
where $n$ is the growth exponent. The presence of one characteristic length
scale during the late times leads to an interesting dynamical scaling
behavior, as can be detected from the density-fluctuation pair-correlation
function $G(\rr,t)=g[\rr/R(t)]$, or the structure factor, $S(\qq,t)=R(t)^dF(\chi)$,
where $d$ is the spatial dimension and $\chi=\qq R(t)$ is the scaled wavevector
~\cite{gunton83}.

Whereas the dynamics of phase separation in alloys, with
conserved order parameter, is quite well understood in terms
of the Lifshitz-Slyozov theory~\cite{lifshitz62},
and is characterized by a growth exponent, $n=1/3$,
independent of spatial
dimension, volume fraction~\cite{gunton83} and even the number of coexisting
phases~\cite{jeppesen93}, the dynamics in fluids is a more complicated phenomenon
due to the coupling of the additional velocity field (which is
absent in alloys) to the ordering field. Consequently,
various competing effects may appear in phase-separating fluids
leading to various growth exponents depending on the strength of the coupling between
the velocity field and the ordering field, on the volume fraction
~\cite{siggia79,furukawa85,sanmiguel85},
on the spatial dimension, and even on the number of components~\cite{laradji94}.

There is no satisfactory theory for the phase separation
dynamics in fluids. Thus our understanding of the phenomenon
is achieved essentially through numerical studies and dimensional analysis
of the relevant dynamical model. 
Using heuristic arguments, Siggia~\cite{siggia79}
was the first to propose 
that the growth exponent is $n=1$ in phase-separating binary fluids
with relatively comparable volume fractions of the two components. This 
growth regime is due to an instability of the tubular domain structure
in binary fluids, leading to the transport of material from the necks to the 
bulges. The numerical studies of the phenomenon are mainly carried 
out by means of 
three different methods: Numerical integration of the corresponding 
kinetic phase-field model known as model H
~\cite{koga91}; 
lattice-Boltzmann (LB) simulations~\cite{alexander93}; and molecular dynamics 
(MD) simulations~\cite{ma92}. 
In contrast to the first two methods, in a molecular dynamics simulation,
the hydrodynamic modes arize naturally from the microscopic
interactions between the molecules subsequent to a quench into the fluid 
phase. There has been some concerns 
with regard to the validity of molecular dynamics
in studying the late-time dynamics of phase separation due to the very small 
time scale involved. It should be noted that phase separation
in simple fluids is naturally a very fast process. Therefore, in order to
probe the dynamics, experimentalists must perform very shallow quenches,
using the advantage of the increased time scale due to critical slowing down.
In contrast, quenches are very deep in a typical molecular dynamics 
simulation.

The numerical integration of model H leads to an asymptotic growth exponent,
$n=1$, in agreement with Siggia's prediction.
LB simulations also find the same result
~\cite{alexander93}.
However, a recent MD simulation on the two-component
Lennard-Jones potential by Ma {\em et al.}~\cite{ma92} suggests
a growth regime with an exponent very close to $2/3$. As we will
see later, such an exponent is due to inertial effects, and can be calculated from
dimensional analysis. A more recent model H simulation by Lookman {\it et al.}
~\cite{lookman96}
finds that by decreasing the shear viscosity of the fluid, a growth exponent
of $n=2/3$ can be observed.
We are therefore faced with the problem that whereas
numerical simulation calculations in the case of phase separation in alloys
agree with the theoretical predictions, numerical simulations which are
expected to most faithfully describe the true dynamics, i.e., molecular
dynamics simulations, are not in agreement with theoretical predictions
in the case of phase separation in binary fluids.
In order to elucidate this apparent
discrepancy between the previous numerical studies and the MD
simulations of Ma {\it et al}, we have carried out a large-scale
and systematic molecular dynamics simulation of the two-component Lennard-Jones
model and found results which disagree with the MD simulation of Ma {\it et al.}
but are fully consistent with experiments and previous model 
H and lattice-Boltzmann simulations.
It is worth noting that the present study is the first large-scale 
MD simulation on three-dimensional binary fluids in which the viscous
regime is observed.

In our simulation model, we consider $N$ monoatomic molecules interacting through the
following two-component Lennard-Jones potential:
\begin{eqnarray} \label{eq:lennard-jones}
U_{\alpha_i,\alpha_j}({\bf r}_{ij})&=&4\epsilon\biggl{\{}\biggl{[}\biggl{(}\frac{\sigma}{r_{ij}}\biggl{)}^{12}
-\biggl{(}\frac{\sigma}{r_{ij}}\biggl{)}^6\biggl{]}\nonumber\\
&-&\biggl{[}\biggl{(}\frac{\sigma}{r^c_{\alpha_i\alpha_j}}\biggl{)}^{12}
-\biggl{(}\frac{\sigma}{r^c_{\alpha_i\alpha_j}}\biggl{)}^6\biggl{]}\biggl{\}}
\theta(r^c_{\alpha_i,\alpha_j}-r_{ij}),
\end{eqnarray}
with $\alpha_i=1$ if $i$ is an A-molecule, and $\alpha_j=2$ if $i$ is a B-molecule.
In Eq.~(\ref{eq:lennard-jones}), $r_{ij}$ is the distance separating the $i^{\rm th}$
molecule
from the $j^{\rm th}$ molecule, and $r^c_{\alpha_i\alpha_j}$ is a cutoff distance
which is equal to $2.5\sigma$ for $\alpha_i=\alpha_j$
and $2^{1/6}\sigma$ for $\alpha_i\neq\alpha_j$. $\theta(x)$ is the standard 
Heaviside function. 
The phase diagram of this model, which has recently been calculated by 
means of mean field theory and Monte Carlo simulation, has a consolute point
at $T_c\sim(4.7\pm0.2)\epsilon$ for a fluid density of $\rho=0.8\sigma^{-3}$
~\cite{toxvaerd95}.
We have performed critical quenches at temperatures ${\rm k_B}T=2$, 3, 3.5, 3.75
and 4$\epsilon$ as well as off-critical quenches at ${\rm k_B}T=2\epsilon$.
Notice that we have not made quenches to very low temperatures in order
to avoid the solid-gas coexistence region. 
The temperature is controlled by a Nos{\'e}-Hoover
thermostat~\cite{nose84}, and the Hamilton equations are integrated using the Leap-Frog 
algorithm with a time step of $\Delta t=0.005\tau$ where the time scale is
$\tau=\sqrt{\mu\sigma^2/\epsilon}$ with $\mu$ being the molecular mass. 
In all of our simulations, the total
number of molecules is $N=343\ 000$, an order of magnitude
larger than the largest 
system size considered by Ma {\it et al.}~\cite{ma92}.
Our simulations were performed on an IBM SP2 parallel machine using 12 processors.
Furthermore, a statistical average is performed for each quench;
16 runs for ${\rm k_B}T=2\epsilon$ and 4 runs for all other quenches.

\begin{figure}
  \psfig{figure=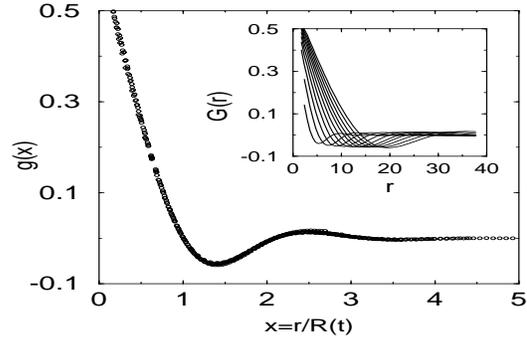,angle=270,width=8cm,height=4.5cm}
  \caption{
Scaled pair-correlation function, $g(x)$, versus the scaled distance,
$x=r/R(t)$, for a quench at ${\rm k_B}T=2\epsilon$. The data shown range
from $t=80$ to $220\tau$. 
The inset shows the time evolution of the correlation function
from $t=20$ to $220\tau$ in steps of $20\tau$.
}
\end{figure}

We have calculated both the correlation function $G({\bf r},t)=
\langle\phi({\bf r},t) \phi(0,t)\rangle$ , where 
$\phi({\bf r},t)=[\rho_A({\bf r},t)-\rho_B({\bf r},t)]/\rho$ is the
order parameter and $\rho_A$ and $\rho_B$ are the local densities of the two components. We have
also calculated
the structure factor $S({\qq},t)=\langle |\tilde \phi(\qq,t)|^2\rangle/V$,
where $\tilde\phi(\qq)$ is the Fourier-mode of the order parameter and $V$ 
is the system volume.
Both the structure factor and the correlation function are then
spherically averaged.
The average domain size is then defined as the first zero of the correlation
function, $R_G(t)$, and as the $n^{\rm th}$ moment of the structure factor,
$R_n(t)=2\pi{[}\int {\rm d}q S(q,t)/\int {\rm d}q q^n S(q,t) {]}^{1/n}$.

The time evolution of the pair-correlation function is shown in the inset of
Fig.~1 for a quench at ${\rm k_B}T=2\epsilon$. The presence of the decaying
oscillations in $G(r,t)$ indicates the occurrence of phase-separated 
domains which are correlated within short distances, due to the conservation
of the ordering field.
The first zero of the correlation
function increases with time implying a coarsening of the
domain structure. We have verified that the system has reached 
a dynamical scaling regime by observing the scaling of the
correlation function, shown in Fig.~1, for times
larger than about $t=80\tau$. 
Very good scaling is also observed in the structure factor (not shown).
The presence of a unique length scale in the system at late times
implies that the width of the interfaces become vanishingly small
compared to the domain size. As a result, the structure factor
should scale as $q^{-(d+1)}$ for large $q$, which is known as Porod's law
and is usually observed in phase-separating systems at late times.
Indeed, we found that the structure factor
is consistent with Porod's law, implying that the phase separation
process in our simulations is well within a dynamical scaling regime.

Now that we are confident that the systems, we are dealing with in our 
simulations, are safely within a scaling regime, we turn to the discussion of
the nature of the growth law. In Fig.~2, the time
dependence of the average domain size, 
as calculated from the various definitions, is shown. Notice the linear dependence of $R(t)$ at late times indicating that
the growth regime should be viscous, in agreement with Siggia's prediction~\cite{siggia79}.
However, when plotting the data in a double-logarithmic plot, we find
that the late-time growth exponent is more consistent with 2/3,
possibly indicating that the observed growth regime
is inertial, as suggested by the MD simulation of Ma {\it et 
al.}~\cite{ma92}. It should be pointed out, however, that the growth law, 
$R(t)=R(0)+a t$, investigated over a finite time range may show a growth exponent which is smaller
than 1 due to a non-negligible value of $R(0)$ and possible other non-algebraic dependences.

In order to determine the true asymptotic growth law, we have to consider the 
relevant dynamical model and analyze our results in light of its implications.
The dynamics of phase separation in fluids can be described by
the so-called model H~\cite{hohenberg77}, 
corresponding to a generalized Cahn-Hilliard 
equation coupled to the Navier-Stokes equation. 
The appropriate dynamical equations can then be written as follows:
\BE \label{eq:cahn}
\partial_t\phi(\rr,t)+\vv\cdot\nabla\phi(\rr,t)
=M\nabla^2\frac{\delta {\cal F}\{\phi\}}{\delta \phi(\rr,t)},
\EE
\begin{eqnarray}\label{eq:navier}
\rho\left[\partial_t\vv(\rr,t)+
\left(\vv(\rr,t)\cdot \nabla\right)\vv(\rr,t)\right]
=\eta\nabla^2\vv(\rr,t)\nonumber\\
-\nabla p(\rr,t)-\phi(\rr,t)
\nabla\frac{\delta {\cal F}\{\phi\}}{\delta\phi(\rr,t)},
\end{eqnarray}
where $\phi(\rr,t)$, $\vv(\rr,t)$ and $p(\rr,t)$ are the local order parameter,
the velocity field and the pressure field, respectively. The constants $M$, $\rho$ and 
$\eta$ correspond to the order parameter mobility, 
the fluid density and the shear viscosity,
respectively. ${\cal F}$ is the usual $\phi^4$ free energy functional~\cite{gunton83}.
The difference between Eq.~(\ref{eq:cahn}) and the usual
Cahn-Hilliard equation is the presence of the second term on the left-hand side
which accounts for the transport of the order parameter by the velocity field. 
Eq.~(\ref{eq:navier}) is different from the usual Navier-Stokes equation
by the presence of the additional force acting on the fluid 
due to gradients in the chemical potential.

\begin{figure}
  \psfig{figure=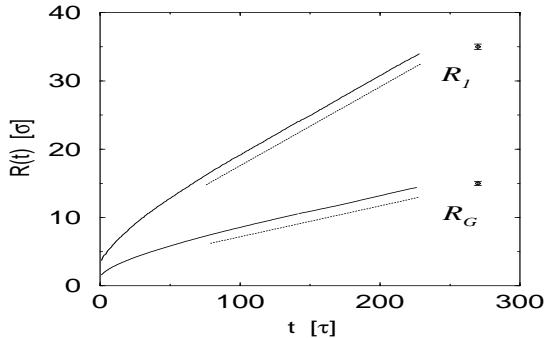,angle=270,width=8cm,height=4.5cm}
  \caption{
The average domain size as a function of time for a quench at ${\rm k_B}T=2\epsilon$.
$R_G(t)$ is the first zero of the pair-correlation function, and $R_1(t)$
is calculated from the first moment of the structure factor. The two dots
indicate the typical size of the error bars in the numerical data, 
and the two dotted lines are straight lines.
}
\end{figure}
The set of equations, (2) and (3), is very difficult to solve, but one can obtain various 
growth regimes by means of simple dimensional analysis. Here we will
limit  ourselves to three dimensions.
At relatively early times, but late enough so that the 
domains are well defined and much larger 
than the interfacial width, the velocity field is decoupled from the
order parameter leading to the usual Lifshitz-Slyozov growth
law usually observed in alloys, $R(t)\sim(M\gamma t)^{1/3}$, where $\gamma$ 
is the interfacial tension~\cite{lifshitz62}. 
This regime will be referred to as the diffusive regime.
At later times, the coupling between $\phi$ and $\vv$ 
cannot be neglected, but the inertial term in Eq.~(\ref{eq:navier}), 
can be neglected, so that $\vv$ becomes slaved by $\phi$. 
One thus obtains the following growth law, $R(t)\sim
(\gamma t/\eta)$, which will be associated with a viscous regime,
and has been predicted by Siggia~\cite{siggia79}
as a consequence of a necking-down instability
of the tubular (interconnected) domain structure due to the transport of
material from the necks to the bulges. This regime has been observed in several
simple binary fluids and binary homopolymer blends
~\cite{wong81,guenoun87} as well as in numerical simulations
~\cite{koga91,alexander93}.
At even later times, the inertial
term in the Navier-Stokes equation can no longer be neglected, 
and one finds the growth law of the inertial regime,
$R(t)\sim (\gamma /\rho)^{1/3}t^{2/3}$~\cite{furukawa85}. 
The two last regimes can be observed
only for interconnected domain structures. For dilute binary solutions,
the domains are droplet-like, and the domain growth is essentially due to
their coalescence leading to a growth law, $R(t)\sim t^{1/3}$, but
with a prefactor which is larger than that in the diffusive regime.
The inertial regime has not been observed experimentally, but it has
been observed in several numerical simulations in two dimensions 
~\cite{alexander93,velasco93,wu95}. Introducing the following two
time scales, $t_v=(\gamma/\eta)t$ and
$t_i=(\gamma/\rho)^{1/3}t^{2/3}$, for the viscous regime and 
the inertial regime respectively,
the $t_v$-dependence ($t_i$-dependence) of $R(t)$ during the viscous
(inertial) regime must be linear and independent of the quench depth,
except maybe for interference with $R(0)$.

\begin{figure}
  \psfig{figure=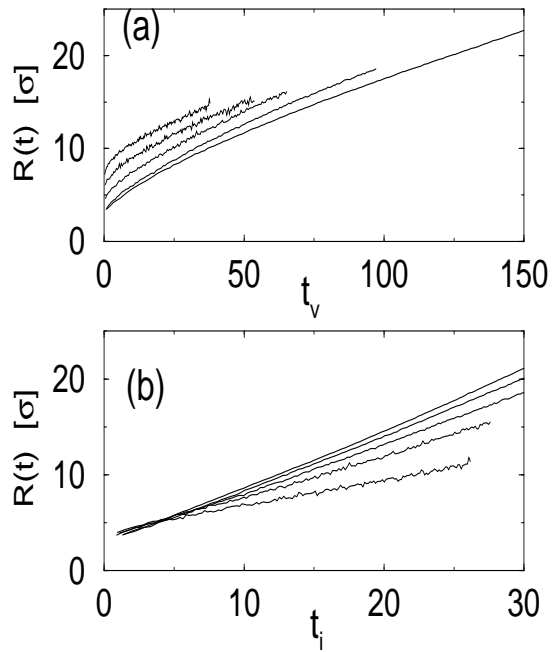,angle=270,width=8cm,height=9.25cm}
  \caption{
(a) The average domain size as a function of $t_v=(\gamma/\eta)t$. Data lines from bottom to top
correspond to ${\rm k_B}T=2$, 3, 3.5, 3.75 and $4\epsilon$ respectively. 
For the sake of clarity, data
data has been shifted vertically upwards.
(b) The average domain size as a function of $t_i=(\gamma/\rho)^{1/3}t^{2/3}$.
Data from top to bottom
correspond to ${\rm k_B}T=2$, 3, 3.5, 3.75 and $4\epsilon$ respectively.
}
\end{figure}

We have therefore calculated, by molecular dynamics simulations,
the interfacial tension, $\gamma$,  and the shear viscosity, $\eta$,
for the various quench temperatures considered in the present study.
We obtain a shear viscosity which is practically independent of temperature,
and equal to $\eta=1.65$. However, the interfacial tension is found to decrease
with temperature, almost linearly, from $(1.85\pm0.07)\epsilon/\sigma^2$ 
for ${\rm k_B}T=2\epsilon$ to $(0.39\pm0.13)\epsilon/\sigma^2$ 
for ${\rm k_B}T=4\epsilon$, since the present model belongs to the
Ising universality class in $d=3$.
In Fig.~3(a), $R(t)$ is plotted versus $t_v=(\gamma t/\eta)$, and 
in Fig.~3(b), $R(t)$ is plotted versus $t_i=(\gamma/\rho)^{1/3}t^{2/3}$
for all quench temperatures. Although the data is almost linear with $t_i$
for all temperatures, the slope of $R(t)$ versus $t_i$ depends strongly
on $T$, whereas the slope of $R(t)$ versus $t_v$ is independent of
temperature. This, therefore, strongly indicates that the growth
regime found in this system cannot be inertial, in contrast to the 
prediction by Ma {\it et al}~\cite{ma92}, 
but in agreement with the other numerical
studies and experiments. 
In our simulations, dynamical scaling is observed starting from $t=80\tau$
at the lowest quench temperatures. At higher temperatures, the 
scaling regime is delayed to later times. This is to be contrasted to the 
study of Ma {\it et al.}, in which it was found
that the scaling regime starts as early as
$20\tau$~\cite{ma92}. 
Of course, the fact that we did not observe an inertial regime does
not disprove the presence of this regime at even later times, as predicted by
the scaling analysis. The inertial regime has been observed 
in previous numerical studies in two dimensions~\cite{velasco93,lookman96},
and in a recent model H simulation in three dimensions~\cite{lookman96}.
In order to detect such a regime, we must simulate
much larger systems.

Another reason, making us even more confident that the dynamical regime
found in the present study is viscous, is the value of the prefactor
of the growth law in terms of $t_v$. 
Siggia predicted that this prefactor is 0.6, whereas San Miguel,
Grant and Gunton~\cite{sanmiguel85}
find that it should be 0.25 from a linear stability 
analysis of the tubular structure. However, a detailed experimental study 
by Guenoun {\it et al.} find that the prefactor is $0.138\pm0.006$~\cite{guenoun87}. 
In our simulation, we find the prefactor to be $0.11\pm 0.01$, which is very 
close to the experimental value of Guenoun {\it et al.}, but disagrees with 
the two theoretical predictions which, however,
are quite crude in nature.
The difference between the value of our prefactor
and that of Guenoun {\it et al.} might be due to the finite size of our systems,
leading to a cutoff of the long-range hydrodynamic modes. Indeed, one expects
that this prefactor decreases linearly with $1/L$ from its thermodynamics-limit
value~\cite{dunweg93}.
Moreover, we found that the prefactor of $t_v$ is independent of volume fraction
for quenches at volume fractions around 0.5.
However, for volume fractions smaller than about
0.3, we found a growth exponent consistent with 1/3. We should notice that 
recently, Nikolayev {\it et al.} have predicted that a sharp transition
from the viscous growth to coalescence-dominated growth occurs at a volume
fraction around 0.3~\cite{nikolayev96}.

In conclusion, we have performed a large-scale systematic molecular dynamics
study of the phase separation dynamics in binary fluids in three dimensions
which faithfully accounts for hydrodynamic modes.
During late times, the system reaches a dynamical scaling regime during which
the average domain size grows linearly with time in agreement with Siggia's
prediction, previous numerical integration of model H, and lattice-Boltzmann 
simulations. The discrepancy with a previous molecular dynamics study has
been clarified.

The authors would like to thank R.C. Desai and P. Padilla for useful discussions.
This work was supported by the Danish
Natural Science Research Council, the Danish Technical Research Council
and by the Natural Sciences and Engineering Research Council of Canada.


\begin{thebibliography}{00}

\bibitem{gunton83} J.D. Gunton, M. San Miguel, and P.S. Sahni,
in {\it Phase Transitions and Critical Phenomena},
edited by C. Domb and J. L. Lebowitz, Vol. 8
(Academic Press, New York, 1983), p.265; and references therein.

\bibitem{lifshitz62} I.L. Lifshitz and V.V. Slyozov,
J. Phys. Chem. Solids {\bf 19}, 35 (1962).

\bibitem{jeppesen93} C. Jeppesen and O.G. Mouritsen,
Phys.  Rev. B {\bf 47}, 14724 (1993).

\bibitem{siggia79} E.D. Siggia,
Phys. Rev. A {\bf 20}, 595 (1979).

\bibitem{furukawa85} H. Furukawa,
Adv. Phys. {\bf 34}, 703 (1985).

\bibitem{sanmiguel85} M. San Miguel, M. Grant, and J.D. Gunton,
Phys. Rev. A {\bf 31}, 1001 (1985).

\bibitem{laradji94} M. Laradji, O.G. Mouritsen, and S. Toxvaerd,
Euro. Phys. Lett. {\bf 28}, 157 (1994); Phys. Rev. E
{\bf 54}, 3673 (1996).

\bibitem{koga91} T. Koga and K. Kawasaki, 
Phys. Rev. A {\bf 44}, R817 (1991);
S. Puri and B. D{\"u}nweg,
Phys. Rev. A {\bf 45}, R6977 (1992);
O.T. Valls and J.E. Farrell,
Phys. Rev. E {\bf 47}, R36 (1993);
A. Shinozaki and Y. Oono,
Phys. Rev. E {\bf 48}, 2622 (1993);

\bibitem{alexander93} J.F. Alexander, S. Chen, and D.W. Grunau,
Phys. Rev. B {\bf 48}, 634 (1993);
S. Chen and T. Lookman,
J. Stat. Phys. {\bf 81}, 223 (1995).


\bibitem{ma92} W.-J. Ma {\it et al.}
Phys. Rev. A {\bf 45}, R5347 (1992).

\bibitem{lookman96} T. Lookman {\it et al.}
Phys. Rev E {\bf 54} (1996).

\bibitem{toxvaerd95} S. Toxvaerd and K. Velasco,
Mol. Phys. {\bf 86}, 845 (1995).

\bibitem{nose84} S. Nos{\'e},
Mol. Phys. {\bf 52}, 255 (1984);
W.G. Hoover, Phys. Rev. A {\bf 31}, 1695 (1985).

\bibitem{hohenberg77} B.I. Hohenberg and P.C. Halperin,
Rev. Mod. Phys. {\bf 49}, 435 (1977).

\bibitem{wong81} N.-C. Wong and C.M. Knobler,
Phys. Rev. A {\bf 24}, 3205 (1981);
F.S. Bates and P. Wiltzius,
J. Chem. Phys. {\bf 91}, 3258 (1989);
T. Izumitani, M. Takenaka, and T. Hashimoto,
J. Chem. Phys. {\bf 92} 3213 (1990).

\bibitem{guenoun87} P. Guenoun, R. Gastaud, F. Perrot, and D. Beysens,
Phys. Rev. A {\bf 36}, 4876 (1987).

\bibitem{velasco93} E. Velasco and T. Toxvaerd,
Phys. Rev. Lett. {\bf 71}, 388 (1993).

\bibitem{wu95} Y. Wu {\it et al.}
Phys. Rev. Lett. {\bf 74}, 3852 (1995).

\bibitem{dunweg93} B. D{\"u}nweg and K. Kremer,
J. Chem. Phys. {\bf 99}, 6983 (1993).

\bibitem{nikolayev96} V.S. Nikolayev, D. Beysens, and P. Guenoun,
Phys. Rev. Lett. {\bf 76}, 3144 (1996).


\end{thebibliography}
\end{document}